\documentclass[prl,preprintnumbers,twocolumn,aps,superscriptaddress,showpacs,floatfix]{revtex4}
\usepackage{graphicx,}
\usepackage{graphicx,dcolumn,bm,epsfig,amsmath,amssymb,textcomp,}
\usepackage{float}
\usepackage[utf8x]{inputenc}

\usepackage{xcolor}
\usepackage{bbm}

\begin{document}

\title {Field excitation in fuzzy dark matter near a strong gravitational wave source}

\author{Shreyansh S. Dave\footnote{shreyanshsd@imsc.res.in}}
\affiliation{The Institute of Mathematical Sciences, Chennai 600113, India}
\author{Sanatan Digal\footnote{digal@imsc.res.in}}
\affiliation{The Institute of Mathematical Sciences, Chennai 600113, India}
\affiliation{Homi Bhabha National Institute, Training School Complex,
Anushakti Nagar, Mumbai 400085, India}

\begin{abstract}
The axion-like particles with ultralight mass ($\sim10^{-22}$eV) can be a 
possible candidate of dark matter, known as the fuzzy dark matter (FDM). These 
particles form Bose-Einstein condensate in the early Universe which can explain 
the dark matter density distribution in galaxies at the present time. We study 
the time evolution of ultralight axion-like field in the near region of a strong 
gravitational wave (GW) source, such as binary black hole merger. We show that 
GWs can lead to the generation of field excitations in a {\it spherical shell} 
about the source that eventually propagate out of the shell to minimize the energy 
density of the field configuration. These excitations are generated toward the 
end of the merger and in some cases even in the ringdown phase of the merger, 
therefore it can provide a qualitatively distinct prediction for changes in the 
GW waveform due to the presence of FDM. This would be helpful in investigating the 
existence of FDM in galaxies.  
\end{abstract}

\maketitle

\noindent

\section{I. Introduction}

The fuzzy dark matter (FDM) is considered to be one of the most promising candidates of dark matter. 
In this model, the field is considered to be ultralight scalar field with mass $m$$\sim$$10^{-22}$eV 
which forms Bose-Einstein condensate (BEC) at the early stage of the Universe \cite{FDM0,SFDM,TRG}. Due 
to a very high occupation number in galactic halos, this field behaves like a classical field \cite{FDM0}. 
It has been shown that the axion-like field (pseudo-goldstone boson) with ultralight mass can be a 
good possible field description for such ultralight particles \cite{axionlike}. Since the mass of the 
particles in this model is very small, the corresponding de Broglie wavelength is of the astrophysical 
scales. Below this scale, the {\it uncertainty principle} leads to the stability of density 
perturbations in FDM against the gravitational collapse \cite{FDM0}. In this way, this model of dark 
matter resolves the overprediction in the number of substructures in dark matter halos and {\it cuspy 
core problem} that arise with the WIMPs model (weakly interacting massive particles model) of dark matter 
\cite{FDM0,SFDM}. 

As the FDM is expected to be present in the form of granular structure (of length scale $\sim$1 kpc) 
inside the galactic halos \cite{natureFDM}, where in the background of this medium other astrophysical 
events, e.g., binary black hole (BBH) mergers, binary neutron star (BNS) mergers, and supernova (SN) 
explosions, could be occurring, and generating a strong gravitational waves (GWs). Therefore, a natural 
question arises that whether such a GW source has any effect on FDM, or whether FDM can affect the source 
so that the produced GWs get modified with observable changes in the waveform. In this regard, the 
following are the phenomena that have been discussed in the literature. The exponential growth of 
occupation number of ultralight scalar bosons near a rotating BH or NS has been extensively studied 
under the phenomenon of superradiance instabilities 
\cite{rev_superradiance,spr1,spr2,spr3,spr4,spr_pulsar1,spr_pulsar2}. This creates a bosonic cloud 
around such objects, which causes their spin-down \cite{rev_superradiance}. In addition, it has been 
argued that the level transitions between quantum states of such bosonic cloud, axions (that constitute 
the bosonic cloud) annihilation to gravitons, and bosenova collapse of the cloud can produce observable 
GW signals \cite{spr5,spr_obsv1}. The first two processes produce long-lasting monochromatic GW signals, 
while the last process produces the signal which is expected to be in the distinct pattern of spikes 
separated by periods of time \cite{spr_obsv1}. Further, under these phenomena, various prospects of 
probing and constraining the ultralight bosons with GWs detection at LIGO and LISA have been proposed
\cite{spr_obsv2,spr_obsv3,spr_obsv4,spr_obsv5,spr_obsv6,spr_obsv7,spr_obsv8,spr_obsv9,spr_obsv10}. 

When such a rotating BH or NS is part of a binary, the presence of bosonic clouds affects the inspiral 
motion of the binary leading to an observable signal in the GW waveform on Earth \cite{ULA_BBH,sukanta}. 
In addition, the possibility of resonance transitions between the growing and decaying modes of the 
cloud make the cloud+binary system much more interesting \cite{sprReso1,sprReso2,sprReso3}. These 
resonances occur when the orbital frequency of binary matches the energy difference between growing and 
decaying modes of the cloud. This depletes the cloud by an amount that depends on the parameters of 
cloud+binary system, causing the generation of a greatly enhanced and short-lived monochromatic GWs
\cite{sprReso1,sprReso2,sprReso3,sprReso4}, which terminate much before the binary merger \cite{sprReso1} 
(the time duration of resonance transition is set by $\Delta \Omega/\gamma$, where $\Delta \Omega$ is 
the resonance bandwidth for transition, and $\gamma$ is the rate of change of orbital frequency of binary
\cite{sprReso3}). This phenomenon can be used to set some observational constraints on the ultralight 
axion-like particles \cite{sprReso4}, e.g., the observations of decay in orbital period of binary can set 
a bound on the decay constant of ultralight axion-like particles \cite{SMohanty}, and on parameters of 
axions in general \cite{massiveScalar}. In presence of a dark matter medium, the dephasing in GW waveform 
has also been studied \cite{numericalSim}. The various prospects for axion searches with Advanced LIGO 
through binary mergers are discussed in detail in \cite{axion_search}; see 
refs.\cite{rev_alp_mrgr1,rev_alp_mrgr2} for review. 

In addition to the above effects of binary merger and FDM on each other, our work in 
ref.\cite{our1} suggests an interesting possibility of generation of field excitations 
in FDM near a strong GW source, which can affect the GW waveform toward the end of 
the merger and in some cases even in the ringdown phase of the merger. This does not 
require superradiance instabilities induced by rotating BH or NS (though, the presence 
of bosonic clouds generated due to superradiance instabilities may have additional 
effects on it). Rather, the generation of field excitations due to the phenomenon 
discussed in ref.\cite{our1} is quite general which only requires a strong GW source 
that produces GWs of significantly large frequency, strain amplitude, and duration.  

In ref.\cite{our1}, it has been shown that a sustained spacetime oscillations can excite 
a complex scalar field in symmetry broken phase through {\it parametric resonance}. The 
spacetime oscillations directly couple to the momentum of field-modes, which leads to the 
generation of field excitation even at frequencies much smaller than the mass of 
longitudinal-modes of the field. In the low frequency regime, mainly transverse excitations 
are generated as the modes corresponding to these excitations have zero mass. The field 
undergoes parametric resonance if appropriate momentum-modes of the field corresponding to 
the frequency of spacetime oscillations are present initially. It also has been shown that 
the finite size effects of system set a lowest frequency cut-off to induce this phenomenon: 
only those spacetime oscillations are able to excite the field whose angular frequency is
$\omega$$\gtrsim$$4\pi/L$, where $L$ is the system size \cite{our1}. This arises due to the 
fact that the momentum-modes having wavelength larger than the upper cut-off 
$\lambda_{_L}$=$L$ cannot grow under the resonance process.

However, in the case of explicit symmetry breaking, in addition to finite size effects, the 
lowest frequency cut-off to excite the field under spacetime oscillations is also set by the 
mass of pseudo-goldstone boson \cite{our1}. In the case of ultralight axion-like field, the 
mass of pseudo-goldstone boson is $m$$\sim$$10^{-22}$eV. Therefore to excite the most dominant 
resonance modes of this field, the frequency of spacetime oscillations must be 
$\omega$$\gtrsim$0.1 {\textmu}Hz or equivalently $f_{_{GW}}$$\gtrsim$10 nHz, where 
$f_{_{GW}}$=$\omega/2\pi$ is the frequency of GWs. This required frequency is easily achieved 
by various GW sources, such as BBH mergers, which can produce GWs with a maximum frequency of 
up to $\sim$1 kHz with significantly large strain amplitude. Further, the ultralight bosons 
can have a wide range of possible masses between $10^{-33}$eV to $10^{-10}$eV 
\cite{string_axiverse,review_ULA}. Therefore for the above maximum frequency produced by the 
binary merger systems, the upper mass cut-off of the field that can be excited by the passing 
of GWs will be $m$$\sim$10$^{-11}$eV. Thus, GWs produced by binary mergers can excite ultralight 
axion-like field having almost all possible mass range, except for the field that has a much 
larger mass. Therefore, this phenomenon is certainly not possible for higher mass QCD axions 
as in such case the required GW frequency would be way beyond the reach of any known GW source 
with significantly large strain amplitude (for a discussion on the QCD axions as a dark matter, 
see refs.\cite{axn1,axn2,axn3}).  

Thus, the restriction in the generation of field excitation in FDM should only arise due to 
finite size effects and small strain amplitude (and of course due to short duration 
of GWs, which we discuss in Sect.IIB(ii)). Indeed, these are the two main competing factors 
which can affect the generation of field excitation at different radial distances $r$ from 
the source. For small $r$, the finite size effects restrict the generation of field excitation 
for frequencies that are smaller than the cut-off frequency at 2-sphere for a given $r$, 
whereas for larger $r$, the strain amplitude of GWs becomes so small that it cannot generate 
field excitation. Therefore, the generation of field excitation in FDM is expected to occur 
only in a certain range of $r$, that is, in a {\it spherical shell} about the GW source. 

The generated field excitations in the shell has higher energy density than the outer 
region of the shell. Therefore, as soon as these excitations arise in spherical shell, 
in order to minimize energy density of the field configuration, the field excitations 
start propagating out of the shell. These excitations are the perturbations in FDM 
(a superfluid medium) in terms of fluid velocity, energy density, and pressure on 
the top of a uniform medium with almost zero fluid velocity. They may have acoustic 
(linear) perturbations as well, which propagate in the fluid with the speed of sound. 
Additionally, as soon as GWs pass completely, the excited field in the shell starts 
rolling back toward the minimum of the effective potential, and oscillates about it. 
These various aspects make this phenomenon even more 
interesting.

In this work, we consider ultralight axion-like field as a field description for FDM. 
Our focus is to study the evolution of this field near a strong GW source. In the first 
part of the study, we show that a sustained spacetime oscillations can generate 
excitations in the field that initially has a uniform field configuration with small 
random fluctuations. These excitations are generated under the phenomenon of parametric 
resonance. Due to the spacetime oscillations, the field configuration mainly goes through 
two processes: first, the process of growth of some specific field-modes following the 
resonance conditions for the given spacetime oscillation frequency, and then the process 
of generation of various other field-modes due to non-linear evolution of the field. 
This, in general, may generate local superflow in FDM at a length scale smaller than 
$f_{_{GW}}^{-1}$ but bigger than the length scale of initial fluctuations. In the second 
part, we consider a model GW waveform constructed using parameters of the observed 
waveform for a BBH merger, and show that the GWs produced by such source can also excite 
the field in a {\it spherical shell} about the source. These excitations arise toward 
the end of the merger and in some cases even in the ringdown phase of the merger, which 
may provide a distinct observable imprints on the measured GWs on Earth. By using a 
sine-Gaussian GW waveform, we also determine the parameter range of the waveform for 
which the resonance growth is possible in FDM. For the study, we solve the Klein-Gordon 
equation of motion for the field as in our case the length scale of evolution of the 
system is much smaller than the Compton wavelength $m^{-1}$ of associated particles; the 
non-relativistic approximation, which gives the Schr\"{o}dinger-Poisson equation of motion 
for FDM, is valid only for length scale larger than $m^{-1}$ \cite{FDM0}.

Recently, it has been shown that the presence of clouds of ultralight axion-like field 
around BHs lead to a suppression in the strain amplitude and frequency of GWs in the 
ringdown phase of the BBH merger \cite{sukanta}. This suppression increases with 
the physical characteristic parameters of the cloud such as mass parameter $\tilde{\mu}
$=$GMm/\hbar c$ and amplitude of the field (where $M$ is the total Arnowitt-Deser-Misner 
mass of gravitational system). For a range of values of such parameters of the cloud, 
the Bayesian analysis gives the suppression in the frequency of GWs between 2.1$-$8.6$\%$
\cite{sukanta}. Our results in the present work provide a possible explanation for such 
suppression due to the clouds. As mentioned earlier, in our case, the GWs passing through 
FDM lead to the generation of field excitations around the source that occurs toward the 
end and in the ringdown phase of the merger. Therefore, this should increase the total 
energy of the field configuration. As the energy of field configuration is increased due 
to GWs, therefore because of energy conservation, the GWs should lose their energy in 
these two phases of the merger, which leads to the suppression in strain amplitude and 
frequency of GWs as observed in ref.\cite{sukanta}. In this way, our study along with the 
study in ref.\cite{sukanta} open a new possibility of indirect detection of FDM and 
constraining such dark matter model by studying the discrepancy in between expected GW 
waveform from a well known GW source and the measured waveform on Earth. In the specific 
case of binary merger systems, this discrepancy must arise toward the end and in the 
ringdown phase of the merger. Thus, it provides a qualitatively distinct predictions for 
changes in the GW waveform compared to other prospects discussed earlier. 

This paper is organized in the following manner. In Sec.II, we provide the field 
equation for FDM minimally coupled to gravity with oscillating spacetime metric. 
Then we consider two cases. In Sec.IIA, we consider a continuous GW waveform and 
show that this leads to the generation of field excitation induced by parametric 
resonance. Then in Sec.IIB, we again consider two cases. In Sec.IIB(i), we consider 
a model GW waveform constructed using parameters of BBH merger, and show that this 
can lead to the generation of excitations in FDM in a {\it spherical shell} about 
the GW source, which are generated toward the end of the merger and in some cases 
even in the ringdown phase of the merger. In Sec.IIB(ii), by using sine-Gaussian 
GW waveform, we determine the {\it resonance growth zone} for FDM, for a parameter 
range of the waveform. Finally, we conclude in Sec.III.

\section{II. Field excitation in FDM}

To study the effects of spacetime oscillations on ultralight axion-like field `$a$', 
we consider time-dependent perturbations in the Minkowski metric, such as 
$g_{\mu \nu}$=$diag \big($$-$1,1$-$$h,1$+$h,1\big)$, where $h$$\equiv$$h(t,z)$ and 
$|h|$$<$1. The action of the field on the spacetime manifold with the given metric 
is \cite{axionlike} 
\begin{equation}
 S=\int d^4x \sqrt{-g}\Big[-\frac{1}{2} F^2 g^{\mu \nu} \partial_{\mu} a~ \partial_{\nu} a 
 - \mu^4 (1-\cos a)\Big],
 \label{eq.Lagrangian}
\end{equation}
where $g$=$det(g_{\mu \nu})$=$-$$\big($1$-$$h^2\big)$, and the parameter $F$ has values 
in the range 10$^{16}$GeV $\lesssim$ $F$ $\lesssim$ 10$^{18}$GeV \cite{axionlike}. Mass 
of the field `$a$' is given by $m$=$\mu^2$/$F$ \cite{axionlike}, where 
$m$$\sim$10$^{-22}$eV is taken in this study. The equation of motion for the field is 
given by \cite{our1,carroll}
\begin{equation}
 \Box a -\frac{1}{F^2}\frac{dV}{da}=0, 
\end{equation}
where the effective potential $V(a)$=$\mu^4 (1$$-$$\cos a)$ and the covariant 
d'Alembertian is given by
\begin{equation}
 \Box a = \frac{1}{\sqrt{-g}}\partial_{\mu}\Big(\sqrt{-g} g^{\mu \nu} \partial_{\nu}a \Big).
\end{equation}
In the expanded form, the field equation becomes 
\begin{equation}
 \begin{split}
  \frac{h_t a_t - h_z a_z}{h^{-1}(1-h^2)} - a_{tt} + \frac{a_{xx}}{1-h} + \frac{a_{yy}}{1+h} 
  + a_{zz} - m^2\sin a=0,
\end{split}
\label{eomg}
\end{equation}
where derivatives are defined as $\xi_{\alpha}$=$\partial \xi / \partial \alpha$,  
$\xi_{\alpha \alpha}$=$\partial^2 \xi / \partial \alpha^2$ for $\xi$$\equiv$$(h,a)$ and 
$\alpha$$\equiv$$(t,x,y,z)$.

\section*{A. With continuous GW waveform}
We consider a continuous GW waveform with $h(t,z)$=$\varepsilon \sin \big(\omega(t$$-$$z)\big)$, 
where $\varepsilon$ and $\omega$ are the time-independent strain amplitude and angular frequency 
of GWs, respectively. We define $f(t,z)$=1$-$$h(t,z)$. With this waveform, the above field 
equation becomes 
\begin{equation}
 \begin{split}
  \frac{\varepsilon^2 \omega \sin \big(2\omega (t-z)\big)}{2f(t,z)f(-t,-z)}(a_t + a_z) - a_{tt} + 
  \frac{a_{xx}}{f(t,z)} + \frac{a_{yy}}{f(-t,-z)} \\+ a_{zz} - m^2\sin a=0.
\end{split}
\end{equation}
This field equation leads to the phenomenon of parametric resonance \cite{our1}, in which some 
specific momentum-modes of the field, following the respective resonance conditions, grow with 
time and lead to the field excitation. 

The above equation can be used for the evolution of field `$a$' around a continuous GW source. 
However, as the coefficient of second-order spatial derivative of field with respect to $z$ is 
trivially one, i.e. non-oscillating, therefore the field-modes corresponding to $z$-axis do not 
grow resonantly. Thus, for simplicity of solving the above equation numerically, we take the 
following assumptions: (i) instead of considering three-dimensional space, we solve this 
equation on a two-dimensional surface by assuming that there is no variation of field along third 
axis (i.e., along $z$-axis), (ii) instead of solving it on a surface of 2-sphere, we solve it on 
a torus geometry, where we take lattice structure as a flat sheet that forms $xy$-plane ($z$=0 
plane) and use periodic boundary conditions. Certainly, the quantitative values of our results 
could be affected due to this choice of boundary conditions, as discussed in some detail in 
ref.\cite{our1}. In the regime where finite size effects dominate, the {\it periodic} boundary 
conditions impose more restrictions in the generation of field excitations under the phenomenon 
of parametric resonance in comparison with the use of {\it fixed} boundary conditions \cite{our1}. 
However, other than this, these boundary conditions do not have any qualitative effect on the 
phenomenon. With these assumptions, the above equation reduces to
\begin{equation}
 \begin{split}
  \frac{\varepsilon^2 \omega \sin (2\omega t)}{2f(t)f(-t)} a_t - a_{tt} + \frac{a_{xx}}{f(t)}
 + \frac{a_{yy}}{f(-t)} - m^2\sin a=0,
\end{split}
\label{eom1}
\end{equation}
where the function $f(t)$=1$-$$\varepsilon \sin (\omega t)$. We define dimensionless 
variables as $t'$=$mt$, $x'$=$mx$, $y'$=$my$, $z'$=$mz$, and $\omega'$=$\omega/m$. With 
these variables, the function $f(t)$ remains unchanged, i.e., $f(t')$=$f(t)$, and the 
equation of motion becomes
\begin{equation}
 \begin{split}
  \frac{\varepsilon^2 \omega' \sin (2\omega' t')}{2 f(t')f(-t')} a_{t'} - a_{t't'} + 
 \frac{a_{x'x'}}{f(t')} + \frac{a_{y'y'}}{f(-t')} - \sin a=0.
\end{split}
\label{eom2}
\end{equation}

We take the mass of the field as $m$=10$^{-22}$eV \cite{axionlike}, and a fix value of GW 
frequency as $f_{_{GW}}$=$\omega/2\pi$=250 Hz \cite{GWbbh} which gives 
$\omega'$=1.05$\times$10$^{10}$. For numerical simulation, the validity of the above 
equation demands that the time step of field evolution $\Delta t'$ should be 
$\Delta t'$$\ll$$ 1/\omega'$. Therefore, we take $\Delta t'$=3$\times$10$^{-12}$ as the 
minimum time step, and lattice spacing as $\Delta x'$=$\Delta y'$=2$\Delta t'$. With this, 
$\Delta t$=2$\times 10^{-5}$ s and $\Delta x$=$\Delta y$=12 km. For simulations, we 
consider $N$=200 lattice points in each spatial direction. Therefore, the total surface 
area of the torus will be $N^2\Delta x$$\times$$\Delta y$=5.76$\times$10$^6$ km$^2$. Note 
that to avoid finite size effects on the field evolution, the condition 
$L$=$N \Delta x$$\geq$$2/f_{_{GW}}$ should be satisfied \cite{our1}, where $L$ is the 
system size in each direction. For the given $f_{_{GW}}$ and $N$, the above condition 
becomes $\Delta x$$\geq$12 km, where the chosen lattice spacing is already at the lowest 
threshold. 

For an estimate of the strain amplitude $\varepsilon$, we adopt the following procedure. 
If we consider a 2-sphere around a GW source having the surface area given above, then 
the radius of that sphere will be $r$=677 km; we denote this radius by $r_0$. (At such a 
distance from the source, spacetime curvature may affect the field evolution and also, 
there may be small deviation from the Newtonian approximation taken below. However, as 
the main focus of this work is to show the effects of spacetime oscillations on the field, 
such as resonance growth etc., which should not be affected due to the overall spacetime 
curvature, therefore for simplicity, we ignore this in our study.) Therefore, at this 
distance from the GW source, the strain amplitude will be given by $\varepsilon$=$h_e 
r_e/r_0$, where $h_e$ is the strain amplitude measured on Earth and $r_e$ is the distance 
between GW source and the Earth. We consider a particular GW source that has $h_e$=1.0$
\times$10$^{-21}$ and $r_e$=410$^{+160}_{-180}$ Mpc \cite{GWbbh}. (These values are taken 
just for an estimate of $\varepsilon$, though the source in ref.\cite{GWbbh} produces a 
time varying GWs, which is considered in the next subsection.) This gives $\varepsilon$=$
0.019^{+0.007}_{-0.009}$.

With the given parameters of spacetime oscillations, we perform numerical simulations to 
solve Eq.\eqref{eom2}. For simulations, we prepare an initial field configuration having 
field values close to the minimum of effective potential with small random fluctuations. 
These field fluctuations are necessary to excite the field as the spacetime oscillations 
only couple to the field through spacetime derivatives; see Eq.\eqref{eom2}. These 
fluctuations are not unphysical, rather, can arise naturally due to thermal and/or quantum 
fluctuations. Since the minimum of effective potential is at $a$=0, therefore fluctuations 
in the field $a(x,y)$ at initial time are considered about $a$=0. For this, we consider the 
initial field configuration $a(x,y)$ which varies randomly on lattice points within the 
range [$-\beta,\beta$].

We now estimate the possible range of $\beta$ for FDM. The BBH merger event, which we have 
considered, has occurred at 410$^{+160}_{-180}$ Mpc distance away from the Earth. Thus, at 
the time of event, the age of the Universe should be 12.46$^{+0.59}_{-0.52}$ Byr. At this stage, 
the temperature of Universe could be roughly the same as at present, which is $T$=2.73 K. 
Therefore, we consider the temperature of Universe at the time of event as the present 
Universe temperature and assume that the temperature of FDM is also the same as that of the 
Universe. (However, it should be noted that the location where we are studying the phenomenon 
of field excitations being close to a strong gravitational field. Therefore, it is possible 
that temperature may be locally higher than the ambient temperature of the Universe, which 
may slightly raise the bound of $\beta$ which we calculate here.) 

The thermal energy at a temperature is given by $T$ (we work in the unit system in which 
$k_B$=$c$=$\hbar$=1). Thus, in natural units, $T$ is the amount of energy available in 
the system to rotate the axion-like field $`a$' uniformly to an angle $\beta$ in a given 
volume $(\Delta x)^3$. Thus, by ignoring the gradient energy at the boundary of such 
rotated configuration arising due to fluctuation, we have the relation 
$(\Delta x)^3 \Delta V(a)$=$T$, where $\Delta V(a)$=$V(\beta)$$-$$V(0)$=$\mu^4(1$$-$$\cos 
\beta)$. By using the series expansion of cosine function for small $\beta$ in this 
relation, we obtain
\begin{equation}
 \beta \approx \frac{1}{\mu^2}\Bigg(\frac{2T}{(\Delta x)^3}
 \Bigg)^{\frac{1}{2}}.
\end{equation}
Putting $\mu^2$=10$^{-8}$ MeV$^2$ for $m$=10$^{-22}$ eV and $F$=10$^{17}$ GeV, and the 
values of $\Delta x$ and $T$ as given above, give $\beta$$\approx$1.5$\times$$10^{-22}$ rad. 
Whereas, for the given range of values of $m$ and $F$, the values of $\beta$ varies between 
$10^{-10}$ rad to $10^{-35}$ rad. 

In simulations, $\beta$ smaller than $10^{-7}$ rad (in orders of magnitude) goes beyond 
the smallest precision of computation for potential energy density. Therefore, we only 
can take $\beta$=$10^{-7}$ rad as the smallest value for the study. However, it should be 
noted that the phenomenon in which we are interested is independent from the choice of the 
initial fluctuations of field (of course, the quantitative values of the outcome depend 
on it). Therefore, in this subsection, where we study a general possibility of parametric 
resonance in FDM, we take $\beta$=$\pi/10$, a greatly large value as a choice, and comment 
on the results for lower values of $\beta$. Whereas, in the next subsection, where we deal 
with BBH merger case, we take $\beta$=$10^{-7}$ rad $-$ the smallest value which can be 
taken for the study $-$ and present the results. In most of the cases, we present results 
in terms of percentage growth in potential energy density due to passing of GWs, which we 
show to be roughly independent from the choice of $\beta$. Therefore, we hope that the 
presented results will be valid even for the actual value of $\beta$ as estimated above.

We utilize the periodic property of the effective potential, i.e., its invariance under 
$a$$\rightarrow$$a+2\pi$, while calculating certain quantities such as the total energy 
density. Following this property of effective potential, we bring back the field within 
the range [$-\pi, \pi$] if it goes beyond this range during the evolution. This ensures 
that the gradient energy is calculated correctly. We numerically solve Eq.\eqref{eom2} on 
torus with the given initial field configuration by using the second-order Leapfrog method 
\cite{leapfrog}. 

In earlier work \cite{our1}, it has been shown that a sustained spacetime oscillations can excite 
an initial field configuration, in which some specific momentum-modes of the field grow following 
the resonance conditions for the given oscillation frequency $\omega$. These field excitations are 
induced by the phenomenon of parametric resonance. Under this phenomenon, at the early stage of 
the generation of field excitation, the whole field configuration acquires some specific 
momentum-modes with growing amplitude with time. However, at later times, the field dynamics 
becomes much more complicated because of non-linearity of the field evolution, due to which various 
other momentum-modes of the field are also generated (modes that are not even related with the
resonance conditions). To show this for the present case, we write the field $a$$\equiv$$a(x,y)$ 
in terms of components $a_1$$\equiv$$a_1(x,y)$ and $a_2$$\equiv$$a_2(x,y)$, where $a_1$=$\cos a$ 
and $a_2$=$\sin a$; here we are representing the angular field $a$ in terms of vector 
$a_1 \hat{i}$+$a_2 \hat{j}$ in the internal space of the field. Furthermore, to understand the 
field configuration at each time, we perform Fourier transform of the components $a_1$ and $a_2$ 
as 
\begin{equation}
  \tilde{a}_i(\vec{k},t) = \frac{1}{\mathcal{A}}\int_{b.c.}d^2\vec{x}~~ a_i(\vec{x},t) e^{i\vec{k}.\vec{x}}~; ~~i=1,2,
\end{equation}
where $\mathcal{A}$ is the total area of the system, {\it b.c.} stands for {\it boundary condition}, 
and $\vec{k}$=$k_x \hat{x}$+$k_y \hat{y}$, $\vec{x}$=$x \hat{x}$+$y \hat{y}$ are the momentum and 
position vectors, respectively. 

\begin{figure}
\includegraphics[width=1.0\linewidth]{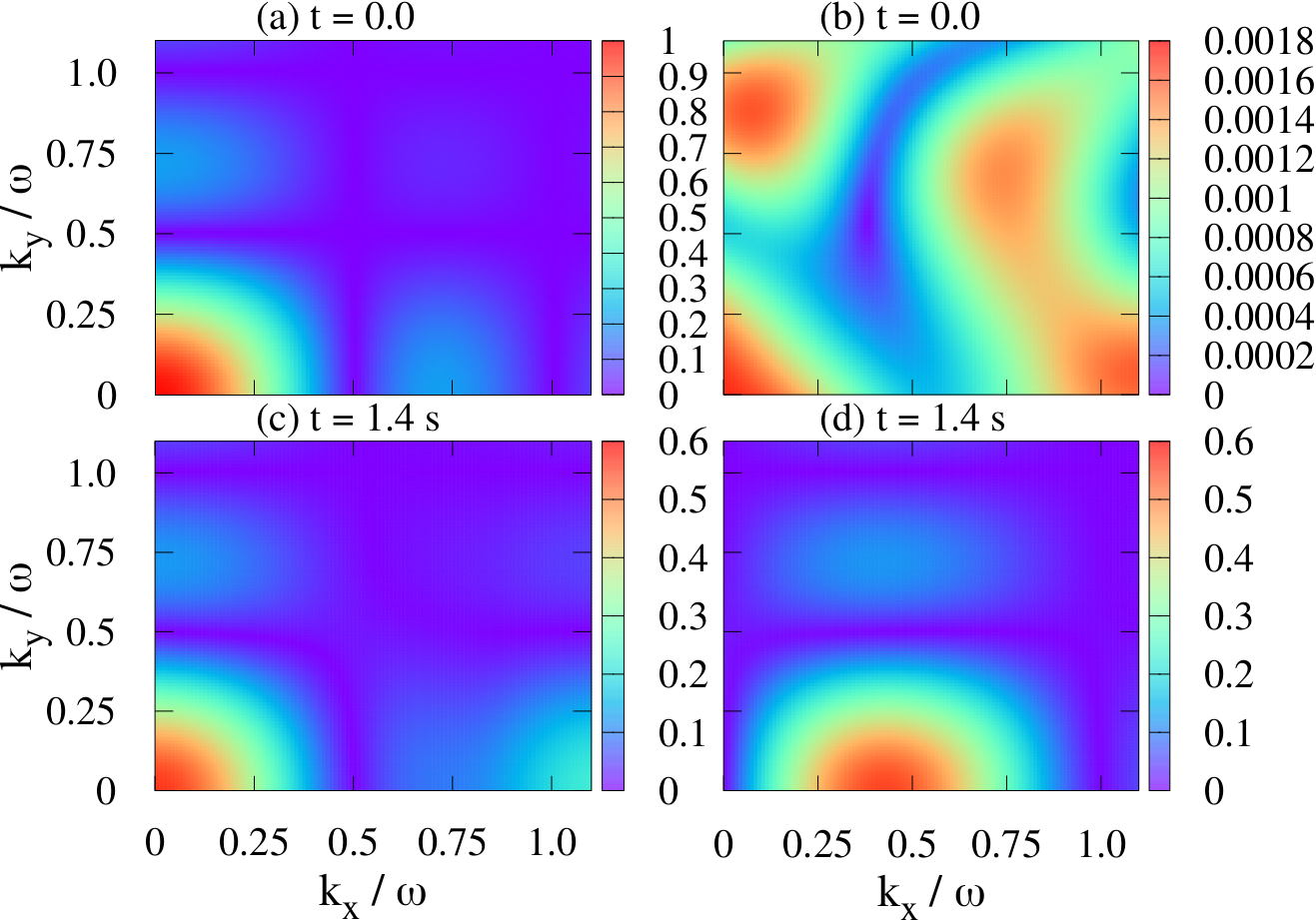}
 \caption{Figure shows the distribution of $|\tilde{a}_1(\vec{k},t)|$ (left panel) and 
 $|\tilde{a}_2(\vec{k},t)|$ (right panel) at initial time and at time $t$=1.4 s. The 
 parameters  of spacetime oscillations are $f_{_{GW}}$=250 Hz and $\varepsilon$=0.026. 
 Figure clearly shows that  under the spacetime oscillations the initial field configuration 
 ($a_1,a_2$) that has dominantly zero-momentum mode is excited to acquire higher 
 momentum-modes following the resonance condition $k_x$$\approx$$\omega/2$, $k_y$$\approx$0.}
 \label{fig1}
\end{figure}

In Fig.\ref{fig1}, we plot modulus of $\tilde{a}_i(\vec{k},t)$ for two time steps of the field 
evolution, at $t$=0 (upper panel) and at $t$=1.4 s (lower panel), in ($k_x/\omega,k_y/\omega$)-plane. 
In the left panel, we plot $|\tilde{a}_1(\vec{k},t)|$, and in the right panel, $|\tilde{a}_2(\vec{k},t)|$. 
For the simulation, the parameters of spacetime oscillations are taken as $f_{_{GW}}$=250 Hz and 
$\varepsilon$=0.026 (maximum possible value of $\varepsilon$ estimated earlier). As mentioned earlier, 
we consider the initial field configuration $a(x,y)$ which varies randomly around zero, so naturally 
the distribution of $|\tilde{a}_1(\vec{k},t)|$ takes a peak at zero-momentum with a value of about 1,
while the distribution of $|\tilde{a}_2(\vec{k},t)|$ shows some random variations depending on the 
initial random fluctuations of the field, with a very small magnitude. This represents the zero-momentum 
BEC state of FDM, which depends on the sequence of random numbers used to generate the initial field 
configuration. At time $t$=1.4 s, due to sustained spacetime oscillations, some specific momentum-modes 
of $a_2$ have grown around $k_x$$\approx$$\omega/2$, $k_y$$\approx$0 ($k_x$$\approx$0.5 peV for the given 
$\omega$) with peak value $\sim$0.6, while the magnitude of zero-momentum mode of $a_1$ has decreased. This 
may generate a superflow in FDM in some local regions that have length scale smaller than $f_{_{GW}}^{-1}$ 
(half-wavelength of the generated field-modes) but bigger than $\Delta x$ (the length scale of initial 
fluctuations) such that the net global flow is zero. We have seen a noticeable growth in these momentum-modes 
starting from the time $t$$\simeq$0.9 s which continues to grow till the time $t$=1.4 s. At later times, with 
further evolution, the field acquires various other momentum-modes because of non-linear field dynamics.  

As mentioned earlier, this growth in the specific momentum-modes is occurring due to  
parametric resonance of the field, where these modes are generated following the 
respective resonance conditions, that is, the relation between $k_x$, $k_y$, $m$, and 
$\omega$. We determine the resonance conditions for Eq.\eqref{eom2} by writing in the 
momentum-space and then linearizing it; see ref.\cite{our1} for more details. We find 
that the resonance conditions for Eq.\eqref{eom2} follow the relation 
$k_x^2$+$k_y^2$+$m^2$$\approx$$(n\omega/2)^2$, where $n$=1,2,3,.... Since in our case 
$\omega$$\gg$$m$, therefore the above relation reduces to 
$k_x^2$+$k_y^2$$\approx$$(n\omega/2)^2$ : an equation of circle in the momentum-space. 
The growth in momentum-modes in Fig.\ref{fig1} clearly validates this relation.

\begin{figure}
\includegraphics[width=1.0\linewidth]{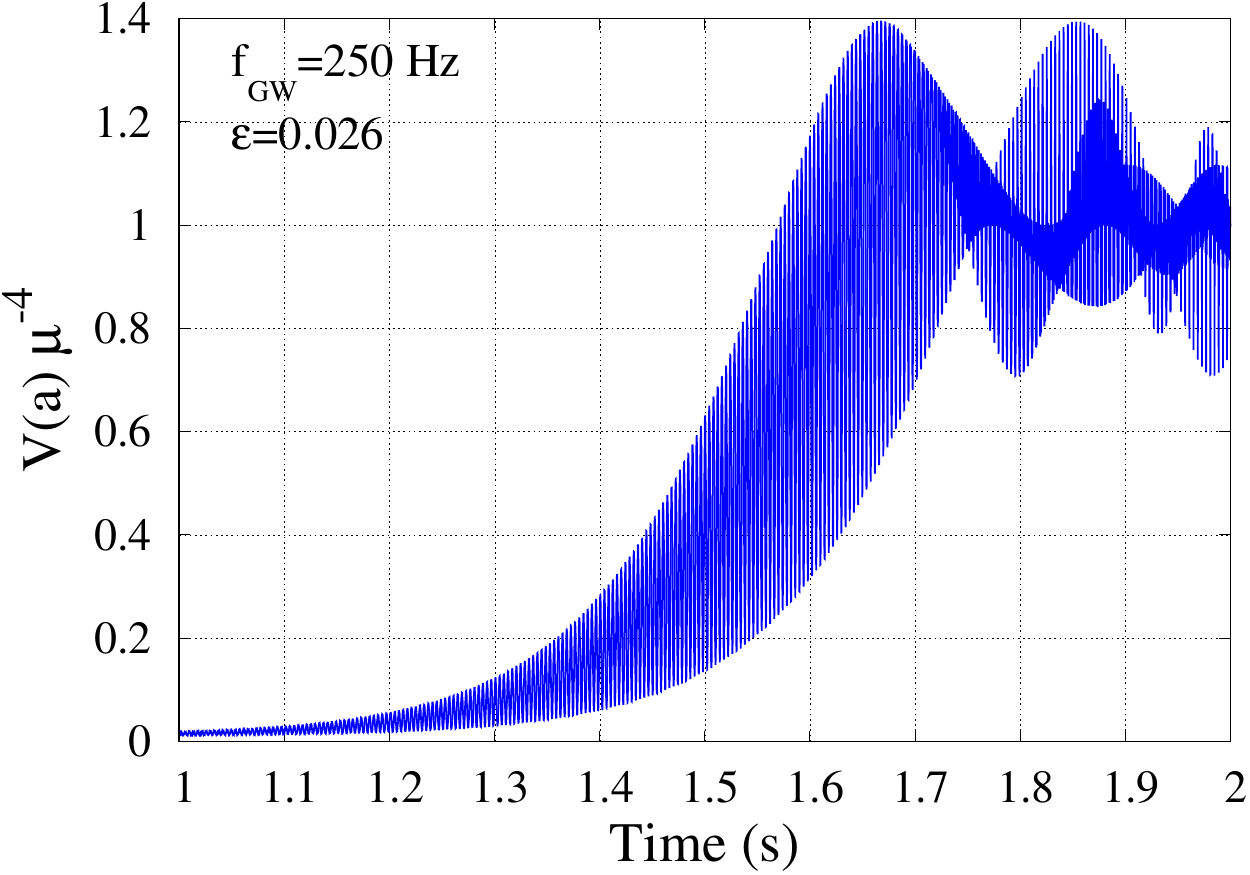}
 \caption{Figure shows the grwoth in average potential energy density (multiplied by $\mu^{-4}$) 
 under the sustained spacetime oscillations.}
 \label{fig2}
\end{figure}

Due to the generation of field excitation, the energy density of the system keeps on increasing 
with time and reaches a maximum value. In simulation, we calculate the total energy density of 
the system $T^{(a)}_{00}$=$\mu^4 T'^{(a)}_{00}$ at each time, where
\begin{equation}
 \begin{split}
  T'^{(a)}_{00}=\frac{a_{t'}^2}{2} + \frac{a_{x'}^2}{2f(t')} + \frac{a_{y'}^2}{2f(-t')} +  (1-\cos a).
\end{split}
\label{eom}
\end{equation}
To calculate these terms, we utilize the periodic property of the effective potential as mentioned 
earlier, under which we bring back the field within the range $[-\pi,\pi]$ if it goes beyond this  
during the evolution. We calculate average values of all above terms on two-dimensional lattice 
at each time. We see the growth in each term of the energy density due to the generation of field 
excitations induced by spacetime oscillations. In Fig.\ref{fig2}, we show the time evolution of 
average potential energy density (multiplied by $\mu^{-4}$) for the given value of $f_{_{GW}}$ 
and $\varepsilon$ (since the scale of the potential energy density is set by $\mu$, we choose to
measure the potential energy density in the units of $\mu^4$). Since in the initial field 
configuration, the field varies randomly between the range $[-\beta,\beta]$ where $\beta$ is a 
small number, the average value of the potential energy density is initially very small. Under the 
sustained spacetime oscillations, when the field achieves full excitations, it covers the whole 
field-space in the range $[-\pi,\pi]$. In such a situation, the average value of $V(a)\mu^{-4}$ 
becomes $\sim$1 as can be seen in the figure.  

We find that, the total energy density of the field achieves roughly two orders of 
magnitude growth from the initial value. As mentioned earlier, for this simulation, 
we have taken $\beta$=$\pi/10$, which gives a large initial fluctuations to the field. 
Instead of it, if we take $\beta$=$\pi/200$, it gives smoother initial fluctuations to 
the field due to which the total energy density of the initial field configuration is 
reduced compared to the former case. We see that even in the latter case, the resonance 
growth of the field begins around the same time as in the former case, that is, at 
time $t$$\simeq$0.9 s. However, due to the lower initial energy density in the latter case, 
the field takes longer time to achieve full excitations; in the former case, it takes 
the time $t$$\simeq$1.7 s, while in the latter case, it takes $t$$\simeq$2.6 s, to achieve 
full excitations.

We also see that increasing values of $f_{_{GW}}$ and $\varepsilon$ reduces the time to 
achieve field excitations (if the appropriate momentum-modes of the field corresponding to 
frequency $\omega$ are present initially). Thus, a stronger GW source can excite the field 
in relatively shorter time. As discussed earlier, as soon as the spacetime oscillations 
stop, the field starts rolling-back toward the minimum of the effective potential, which 
leads to the oscillations in the field about the minimum.

\section*{B. With time varying GW waveform}

\subsection{(i) BBH merger GW waveform}

In the previous case, we have taken a continuous GW waveform that enables the field 
`$a$' to excite under the phenomenon of parametric resonance. However, BBH and BNS 
mergers do not produce a continuous GWs. Rather, these sources produce a time varying 
GWs whose amplitude and frequency increase until the merger is completed. Therefore, 
the phenomenon of generation of field excitation may become much more complicated in 
this case. In this subsection, a time varying GW waveform is considered to study the 
field excitations near a strong GW source. For simplicity, the GW waveform is taken 
to be
\begin{equation}
\begin{split}
 h_e = 10^{-21}[h_a e^{h_b t}\sin(\omega t)],~~~~~~~~~ t\leq t_0,\\
 h_e = 10^{-21}[h_c e^{-h_d (t-t_0)}\sin(\omega t)],~~ t > t_0,
  \end{split}
  \label{funche}  
\end{equation}
where $\omega$=$2\pi f_{_{GW}}$, $f_{_{GW}}$=$\exp(f_a e^{f_bt}$+$f_c)$, and 
($h_a$, $h_b$, $h_c$, $h_d$, $f_a$, $f_b$, $f_c$, $t_0$) are the phenomenological 
parameters to be fitted using a known GW waveform. (Note that in general, the 
strain amplitude of GWs produced by BBH or BNS mergers are not spherically symmetry 
about the source \cite{waveform1}. However, for simplicity here we ignore this, and 
proceed by considering a spherically symmetric GW strain amplitude about the source. 
We also ignore the cross-polarization component of GWs.) We determine parameters of 
the above function such that $h_e$ shows a qualitatively similar variation as the 
strain amplitude for a known GW source. For this, we consider the GW source GW150914 
\cite{GWbbh}, and model the waveform with the above function. By taking some limiting 
values of $h_e$ from the actual waveform, we obtain $f_a$=2.9$\times$10$^{-8}$, 
$f_b$=42.5 s$^{-1}$, $f_c$=3.465, and the remaining parameters, for the two possible 
values ​​of the merger time $t_0$, as given in the table-\ref{tab:1}.
\begin{table}[ht]
\centering
\begin{tabular}{c c c c c c c}
\hline
  & & $h_a$ & $h_b$(s$^{-1}$) & $h_c$ & $h_d$(s$^{-1}$) & $t_0$(s) \\ [0.5ex]
\hline
 Set-I & & 7.7$\times 10^{-4}$ & 17.5 & 1.3 & 300.0 & 0.425 \\ 
 \hline
 Set-II & & 1.1$\times 10^{-3}$ & 16.4 & 1.3 & 500.0 & 0.43 \\ [1ex]
 \hline
\end{tabular}
\caption{Table provides values of some parameters of GW waveform, given in 
Eq.\eqref{funche}, for two possible values of the merger time $t_0$.}
\label{tab:1}
\end{table}
The values of $h_d$ are taken such that $h_e$ becomes close to zero at time 
$t$=0.44 s \cite{GWbbh} for both the cases. With this, the ringdown phase 
of the merger lasts longer for Set-I parameters than for Set-II parameters.

The GW waveform given by the above function is not exactly the same as 
that produced by GW150914, though the qualitative variation in the strain 
amplitude and frequency of the waveform is similar to this source \cite{GWbbh}. 
It should be noted that the focus of the present study is to demonstrate a 
general possibility of generation of field excitation due to passing of 
GWs, without focusing on any particular GW source. Certainly, a detailed 
structure of the GW waveform can affect the generation of field excitations 
quantitatively. In fact, we show that our results are highly sensitive to 
the parameters of the GW waveform used for the simulation.

To perform simulations, in this case also, we continue with the assumptions 
that there is no variation of the field along $z$-direction (along $r$ in 
spherical polar coordinates) and the consideration of the lattice structure 
as a flat sheet forming $z$=0 plane with periodic boundary conditions. Under 
these assumptions and in terms of dimensionless variables, Eq.\eqref{eomg} 
becomes
\begin{equation}
  \frac{h_{t'} a_{t'}}{h^{-1}(1-h^2)} - a_{t't'} + \frac{a_{x'x'}}{1-h} + 
  \frac{a_{y'y'}}{1+h} - \sin a=0,
\label{eomg2}
\end{equation}
where $h$=$h_e r_e/r$, and from Eq.\eqref{funche},
\begin{equation}
\begin{split}
h_{t'}=\mathcal{R}h_a e^{h_b' t'}\big[h_b'\sin(\omega' t')+ \Omega \big],~~~~~~~~~~~~ t'\leq t_0',\\
h_{t'}=\mathcal{R}h_c e^{-h_d' (t'-t_0')}\big[-h_d'\sin(\omega' t')+ \Omega \big],~ t' > t_0',
\end{split}
\end{equation}
where $\mathcal{R}$=10$^{-21}r_e/r$, $h_b'$=$h_b/m$, $h_d'$=$h_d/m$, $t_0'$=$mt_0$, 
$\Omega$=$(\omega'$+$\omega'_{t'} t') \cos(\omega' t')$,
$\omega'_{t'}$=$\omega' f_a f_b' e^{f_b' t'}$, and $f_b'$=$f_b/m$. Using these 
parameters of GWs, we perform simulations to solve Eq.\eqref{eomg2}. We take the 
same lattice spacing $\Delta x$ (that gives $r$=$r_0$) and time step $\Delta t$ 
as used before, and $\beta$=$10^{-7}$ rad for the initial field fluctuations. 
We take the maximum possible value of $r_e$ for the source GW150914, which is 
$r_e$=570 Mpc \cite{GWbbh}. We follow the same procedure of calculating $r$ and 
strain amplitude $h$ as used in the last subsection. 

\begin{figure}
\includegraphics[width=0.68\linewidth, angle=270]{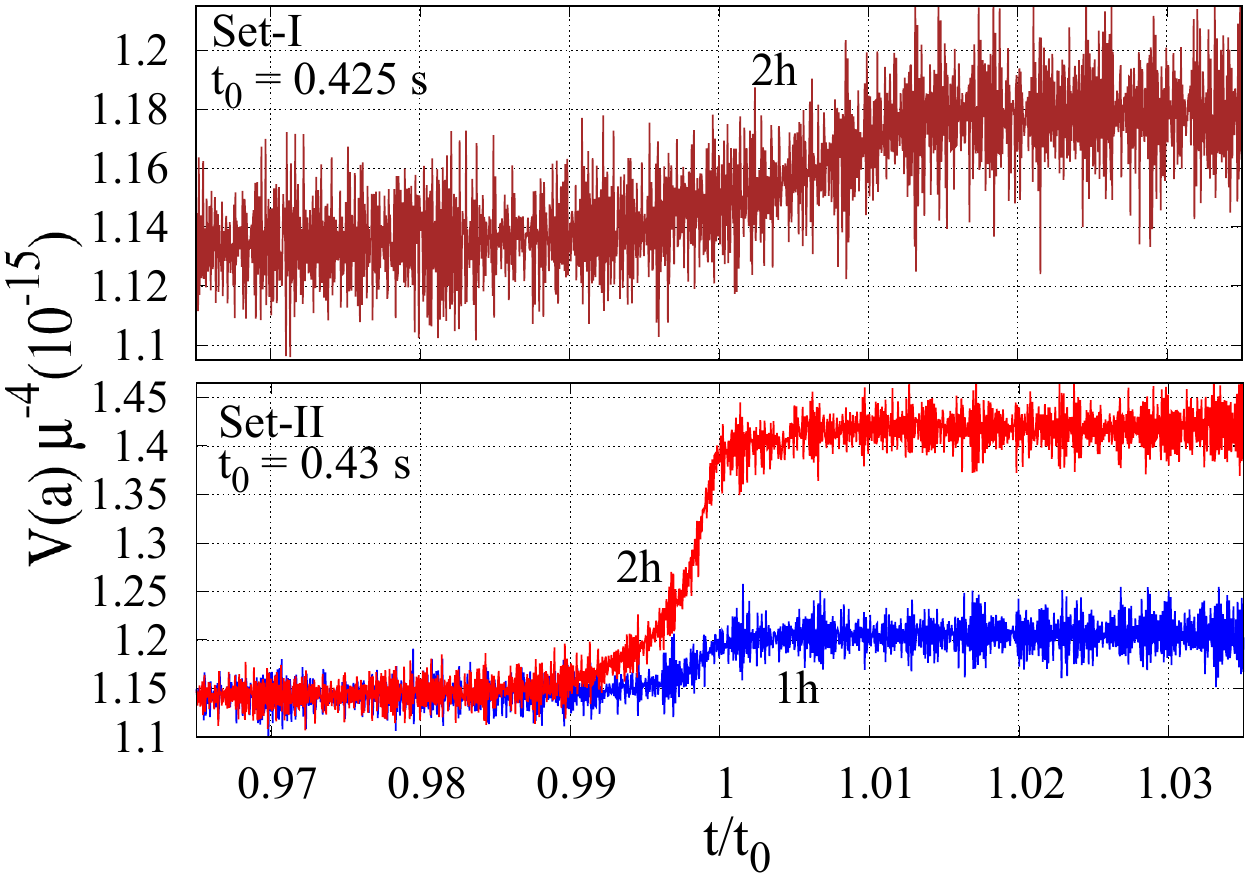}
 \caption{Figure shows the growth in average potential energy density (multiplied by $\mu^{-4}$) 
 at $r$=$r_0$ hypersurface for Set-I (upper panel) and Set-II (lower panel). These growths arise 
 around $t_0$ time, that is, toward the end and in the ringdown phase of the merger. 
 Curve $1h$ corresponds to the case when the actual $h$ has been taken, while curve $2h$ 
 corresponds to the case when factor two is multiplied into $h$.} 
 \label{fig3}
\end{figure}

In Fig.\ref{fig3}, we plot the time evolution of average potential energy density (multiplied 
by $\mu^{-4}$) for the given sets of parameters. Curve $1h$ corresponds to the case when the 
actual $h$ has been taken (as defined in Eq.\eqref{funche}), while curve $2h$ corresponds to 
the case when factor two is multiplied into $h$. In the upper panel, the plot shows the time 
evolution of the potential energy density for Set-I GW waveform, which is only plotted for 
the case of $2h$ (brown). For $1h$, the overall growth in potential energy density is approximately 
1.1$\%$ from the initial value, which is very small to show in the plot. The growth in the 
potential energy density for the case of $2h$ is approximately 3.9$\%$. In the lower panel, the 
plot shows the time evolution of the potential energy density for Set-II GW waveform for $1h$ 
(blue) and $2h$ (red) cases. In these two cases, there are approximately 5.6$\%$ and 24.3$\%$ 
overall growth in the potential energy density from the initial value at the given constant 
$r$-hypersurface. The curve $2h$ in the plot indicates that a relatively strong GW source can 
cause a very large growth in the energy density of FDM. The percentage growth in total energy 
density is also in the same order as that for potential energy density.

The growth in the potential energy density for Set-I and Set-II has basically two distinct features: 
(i) the overall growth in the potential energy density, and (ii) the time-domain of the occurrence 
of these growths. The overall growth for Set-I is lesser than the growth for Set-II. This is because 
for Set-II, a large strain amplitude and frequency are simultaneously present in the waveform toward 
the end of the merger to generate large excitations in FDM, which is somewhat absent for Set-I. On the 
other hand, for Set-I, since the ringdown phase persists for a longer time, the growth in potential 
energy density continues even after time $t_0$ [$0.98t_0 \lesssim t \lesssim 1.013t_0$], which is 
quite different from the case of Set-II in which growth is very rapid around time $t_0$ 
[$0.98t_0 \lesssim t \lesssim t_0$] and stops quickly after it. All this suggests that the generation 
of field excitations is highly sensitive to the parameters of the GW waveform, which based on these 
parameters can be generated toward the end as well as in the ringdown phase of the merger.

We also study the distribution of field-modes in ($k_x,k_y$)-plane at $t_0$ time, as done 
in Fig.\ref{fig1}. We find that for each case in Fig.\ref{fig3}, the initial distribution 
of $|\tilde{a}_2(\vec{k},t)|$ is redistributed at time $t_0$ without much growth, and takes 
peak at a higher momentum causing growth in energy density of FDM.  

We have verified that the percentage growths discussed in Fig.\ref{fig3} are independent from 
the strength of the initial field fluctuations, that is, from the choice of $\beta$. We find 
that for the range of $\beta$ from $10^{-1}$ to $10^{-7}$ rad, the percentage growth in potential 
energy density for Set-II parameters (with $1h$) {\it randomly} varies between 5.1$-$5.7$\%$
(an extremely small {\it random} variation for many orders of magnitude change in $\beta$), 
which shows that it is almost independent from the choice of $\beta$. It indicates that our 
results could be valid even for the actual value of $\beta$, i.e. for $\beta$$\sim$10$^{-22}$ 
rad, as estimated previously for FDM.

It should be noted that due to finite size effects, the field cannot achieve excitations 
until the frequency of GWs becomes equal or exceeds the lowest frequency cut-off 
$f_{_{GW}}$=250 Hz for this hypersurface. Certainly, for higher $r$-hypersurfaces, the 
frequency cut-off will be lesser than $f_{_{GW}}$=250 Hz, which would provide a larger 
time-domain for the field to achieve excitations. However with increasing $r$, the strain 
amplitude also decreases, which can suppress the generation of field excitation. On the 
other hand, although the strain amplitude becomes larger for smaller $r$, the finite size 
effects suppress the generation of excitations. Therefore, the field excitations are 
expected to be generated in a {\it spherical shell} about the GW source with maximum 
excitation at an optimum radial distance $r$. Although this investigation requires full 
(3+1)-dimensional simulations, in the given simulation setup, we now show that our this 
expectation is true. 

\begin{figure}
\includegraphics[width=1.05\linewidth]{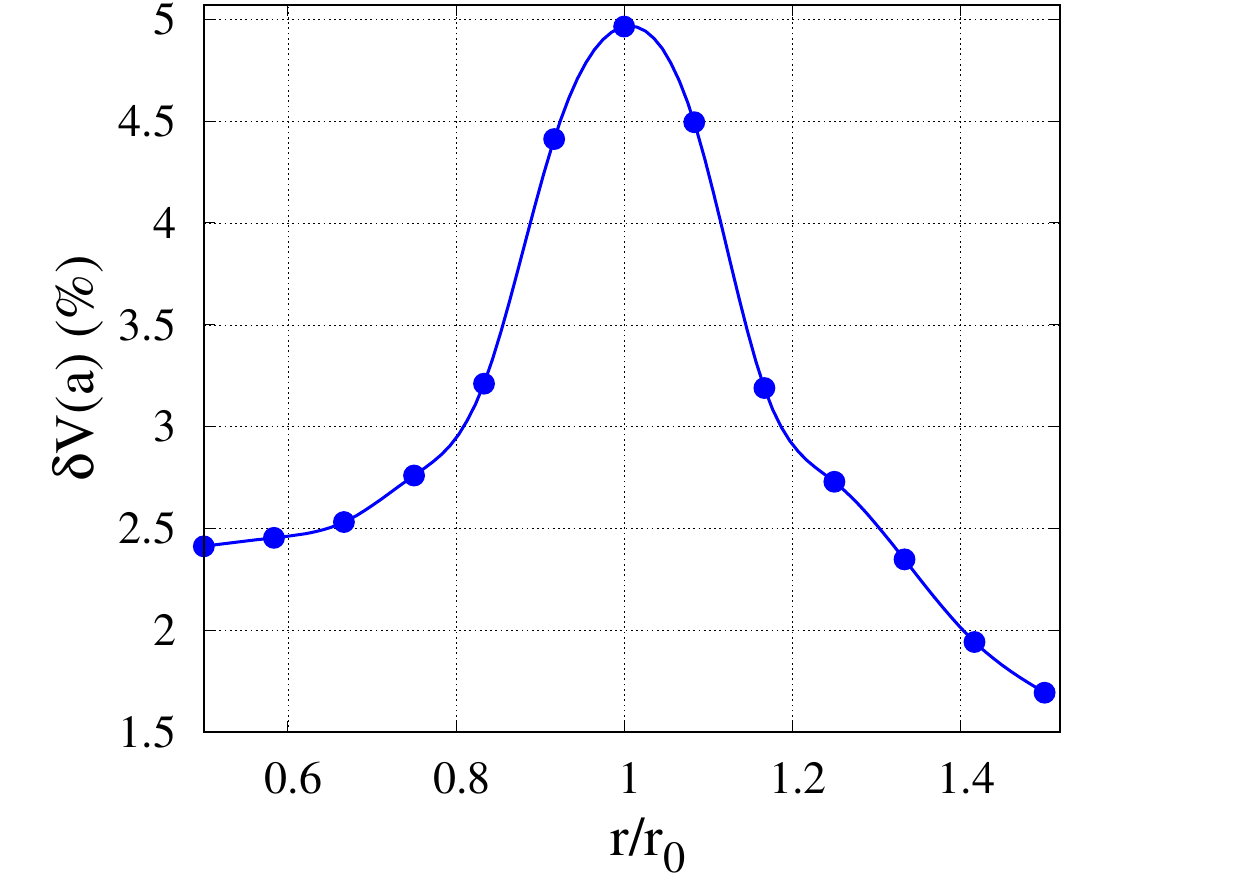}
 \caption{Figure shows the percentage growth in average potential energy density 
 $\delta V(a)$ at different $r$ from GW source; $r_0$=677 km. This entire curve 
 is a result of a combination of finite size effects and strain amplitude, which 
 shows that only in a {\it spherical shell} about the GW source, field excitations 
 can be generated.} 
 \label{fig4}
\end{figure}

In Fig.\ref{fig4}, we plot the percentage growth in average potential energy density 
$\delta V(a)$=$\frac{V(t>t_0)-V(t=0)}{V(t=0)}$$\times$100$\%$ at different $r$. The 
simulations are performed using parameters of Set-II. However, for this study, we stop 
the spacetime oscillations after time $t_0$ for each $r$ to avoid any ambiguity in the 
choice of $h_d$ that may arise while simulating hypersurfaces for $r$$<$$r_0$. This 
consideration does not affect the physical aspects in which we are interested. To change 
$r$ in the simulations, we change $\Delta x$ (=$\Delta y$). As the maximum momentum-mode 
present in the initial field configuration is given by $k_{max}$=$2\pi/\Delta x$, 
therefore with the change in $\Delta x$, the momentum-modes present in the initial field 
configuration also change. This may affect the resonance process quantitatively. However, 
its effects should not be very significant as the overall change in $k_{max}$ is only 
between 69.8 to 209.3 peV for the given range of $\Delta x$ (6.0 $-$ 18.0 km) in 
Fig.\ref{fig4}. As discussed above, the entire curve in Fig.\ref{fig4} is a result of a 
combination of finite size effects and strain amplitude, which clearly shows that the 
field excitations can arise only in a {\it spherical shell} about the GW source.  

Throughout the evolution, the field remains in the linear regime, and hence the 
momentum-modes of the field corresponding to $z$-axis will not be affected by 
the growth of $k_x$ and $k_y$ modes of the field. Therefore, in this case, our 
(2+1)-dimensional simulation is a reasonable approximation to the full 
(3+1)-dimensional simulations. However, as mentioned earlier, the generation of 
field excitation in FDM should cause a loss in GW energy due to energy conservation. 
Therefore, this process of generation of field excitation should modify the actual 
waveform produced by the GW source, especially toward the end of the merger. 
Therefore, to study the back-reaction of the generated field excitations, in 
determining the actual changes in the GW waveform, one needs to solve full 
Einstein's equation, which is not our focus in this work. 

In this study, the results are presented using a model GW waveform that is suitable for an 
observed GW waveform on Earth. We have considered it as an actual waveform produced by a
source, and performed the simulations. With the simulation results, we then conclude that 
the GWs can generate field excitations in FDM around the source, which would modify the GW 
waveform due to energy loss. Thus, this study suggests that the GWs observed on Earth may 
have already been affected due to the presence of FDM. A detailed study will reveal
whether such a change in the GW waveform can be observed experimentally, and if it does, 
it will indirectly show the existence of FDM in the intermediate region between the GW 
source and the Earth.

\subsection{(ii) Sine-Gaussian GW waveform}

In the previous subsection, we have shown, by using a specific GW source, 
that a time varying GWs can generate field excitations in FDM. Indeed, 
some GW sources, such as core-collapse supernovae (CCSNe), generate 
transient GWs whose duration may vary from one source to another. We now 
show that the duration of GWs may affect the generation of field excitations 
in FDM. To analyze it, we consider a sine-Gaussian function as an {\it ad 
hoc} GW waveform \cite{sg1,sg2,sg3,sg4,sg5}, which is also used to model 
the GW waveform produced by CCSNe \cite{sg4,sg5}. This waveform is given by 
\begin{equation}
 h = h_0 \exp(-(t-t_0)^2/\tau^2)\sin(2\pi f_{_{GW}} t),
\end{equation}
where $h_0$ is an amplitude scale factor, $t_0$ is the arrival time of 
GW signal, $\tau$ quantifies the duration of signal, and $f_{_{GW}}$ is 
the frequency of GWs. For simplicity, we again ignore the 
cross-polarization component of GWs. 

In this case also, we use the same simulation parameters as used 
in the last subsections to solve Eq.\eqref{eomg2}. For the study, 
we take $t_0$=0.3 s, and $f_{_{GW}}$=$\{250,500\}$ Hz. The value 
of $t_0$ is chosen sufficiently large compared to $\tau$ so that 
its value cannot affect the growth in the average potential energy 
density. Our focus is to determine the parameter range of this 
waveform for which FDM can be excited. A GW waveform with arbitrarily 
small $\tau$ cannot generate resonance growth in FDM. The same is 
also true for a waveform with arbitrarily small $h_0$. Therefore, 
in the parameter space $(h_0,\tau)$, for each frequency $f_{_{GW}}$, 
there should be {\it resonance growth zone} and {\it growth 
forbidden zone} partitioned by a curve whose horizontal and vertical 
asymptotes are at $h_0$$\neq$0 and $\tau$$\neq$0, respectively. As 
already discussed, the frequency cutoff to generate resonance growth 
is set by the finite size effects on the respective $r$-hypersurface 
from the source. 

\begin{figure}
\includegraphics[width=1.05\linewidth]{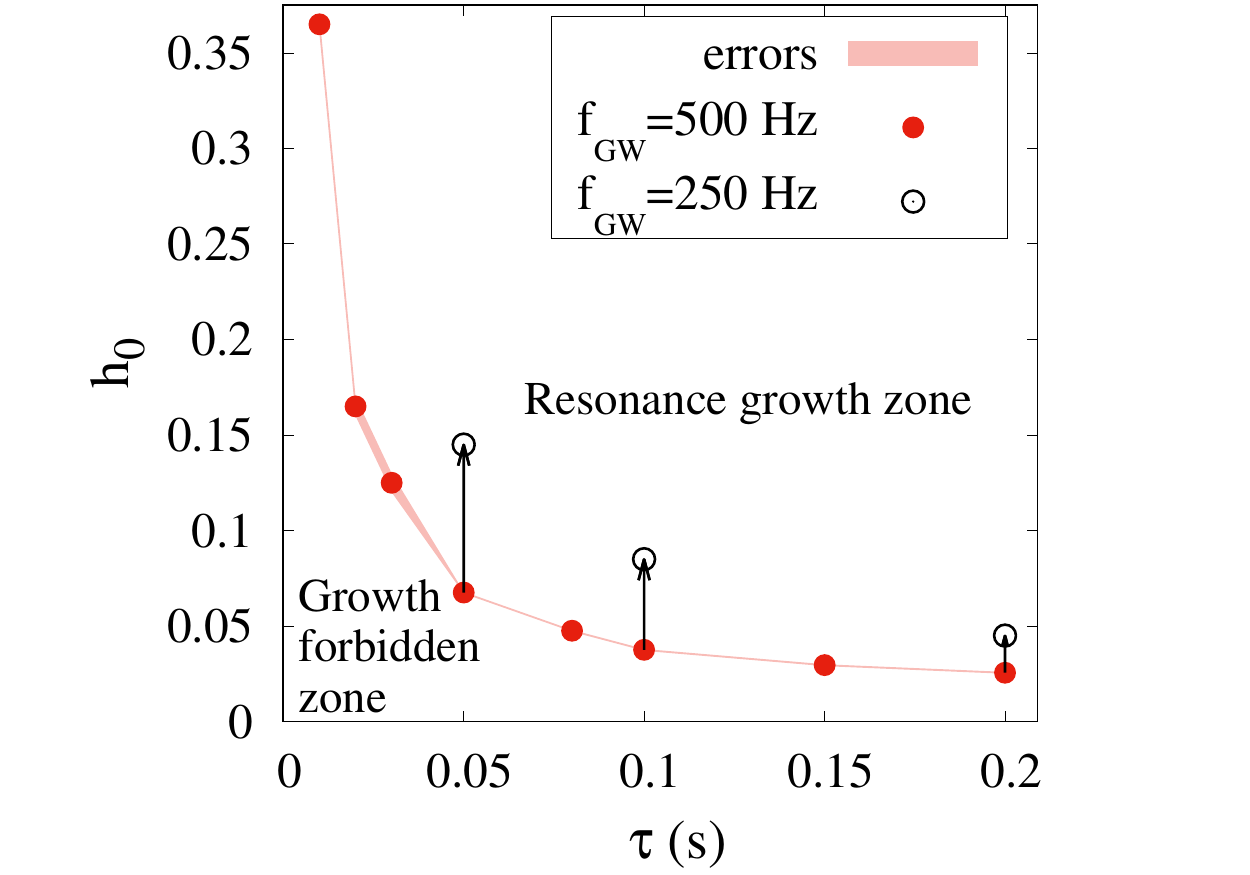}
 \caption{Figure shows the {\it resonance growth zone} and {\it 
 growth forbidden zone} for FDM in parameter space ($h_0,\tau$). 
 The analysis is done for $f_{_{GW}}$=250 Hz (empty circles) and 
 $f_{_{GW}}$=500 Hz (filled circles), at $r_0$ hypersurface; 
 $r_0$=677 km. The {\it resonance growth zone} covers the region 
 above the plotted points for the respective $f_{_{GW}}$. This 
 zone is strongly reduced toward smaller $\tau$ by decreasing 
 frequency, which is depicted by arrows. The errors in the plot 
 indicate the uncertainty in determination of value of $h_0$ for 
 which there is no resonance growth.}
 \label{fig5}
\end{figure}

In Fig.\ref{fig5}, we plot the {\it resonance cutoff} points in 
parameter space ($h_0,\tau$) for each $f_{_{GW}}$, which divide 
the parameter region of possible resonance growth to the {\it 
growth forbidden zone}. In the simulations, we observe resonance 
growth only for those $h_0$ that have value higher than the 
plotted points for each $\tau$. Thus, the {\it resonance growth 
zone} covers the region {\it above} these points for the respective 
$f_{_{GW}}$. With increasing values of $h_0$ and $\tau$ above 
theses points, the overall growth in average potential energy 
density increases. The errors in the plot indicate the uncertainty 
in determination of value of $h_0$ for which there is no resonance 
growth. This plot clearly indicates that the generation of field 
excitations in FDM is only possible with the GWs that have a 
significant duration and strain amplitude. It is also clear from 
Fig.\ref{fig5} that the {\it resonance growth zone} is strongly 
reduced toward smaller $\tau$ by decreasing frequency, which is 
depicted by arrows in the plot. The plotted points for 
$f_{_{GW}}$=500 Hz can be best fitted by the function 
$h_0$=$10^{-3}\Big(5.3\big(\tau$$-$3.3$\times$$10^{-3}$$\big)^{-0.84}$+2.0$\Big)$ 
indicating to have the lowest cutoff for $\tau$ as well as $h_0$ 
to generate resonance growth at the two extremes, $\tau$$\rightarrow$0 
and $\tau$$\rightarrow$$\infty$, respectively.     

The result presented in Fig.\ref{fig5} has important phenomenological
implications for CCSNe. It may be possible that some CCSNe generate 
GWs whose parameters are such that they are capable of generating field 
excitations in FDM in a {\it spherical shell}. It is expected that the 
GWs produced by a CCSN carry the information of explosion mechanism, 
more specifically, the magnitude and character of deviation of the core 
from its spherical symmetry \cite{sg4}; for a review, see \cite{CCSN1}. 
Thus, with the detection of the GWs produced by CCSNe, one is expected 
to be able to probe the dynamics of deep inside the core of SNe. However, 
as we argue that GWs produced by CCSNe, in some parameter range, could 
excite FDM present in the outer space, hence should lose some energies 
due to energy conservation. Therefore, this process may hide some 
relevant information of CCSNe that are expected to be studied by 
detecting GWs. We will explore this in more detail in our future work.

\section{III. Conclusions}

In this work, we have shown that strong gravitational waves (GWs) can generate field 
excitations in fuzzy dark matter (FDM) in {\it spherical shell} about the GW source.
The field description for FDM is considered as an ultralight axion-like field having 
mass $m$$\sim$10$^{-22}$eV, where initially the field is taken at the minimum of 
effective potential with small random fluctuations. These generated excitations can 
lead to the generation of local superflow in FDM. As soon as these excitations arise, 
in order to minimize the energy density of the field configuration, they start 
propagating out of the shell as the perturbations in FDM. These excitations may have 
acoustic (linear) perturbations as well, which propagate with the speed of sound in 
medium. 

We have shown, by using a model GW waveform suitable for a binary black hole (BBH) 
merger, that GWs can generate field excitations in FDM toward the end and in the 
ringdown phase of the merger. Due to energy conservation, while generating these 
excitations, GWs will lose their energy in these two phases of the merger, due to 
which GW waveform should get modified. Thus, our results can provide a qualitatively 
distinct prediction for the presence of FDM around the GW source than the other
prospects discussed earlier in literature. This may be observed in the measurement 
of GWs on Earth, which requires a detailed investigation. A detailed study in 
ref.\cite{sukanta} shows a suppression in the strain amplitude and frequency of GWs 
in the ringdown phase of BBH merger due to the presence of clouds of ultralight 
axion-like field. A possible explanation for such suppression can be given due to 
the phenomenon discussed in the present work. 

The generation of field excitations is not only limited to the above choice of 
parameters of field and GW source, rather, can be possible for a wide range of 
masses of axion-like field and for any strong GW source. Indeed, we have also 
shown that a sine-Gaussian GW waveform can generate field excitations in FDM, 
and determined the {\it resonance growth zone} for a parameter range of this 
waveform. Thus, the main criteria, we obtain to generate the field excitations 
in FDM, are: (i) the angular frequency of GWs should be $\omega$$\geq$$m$, (ii) 
the surface area of the 2-sphere about the source must be large enough to avoid 
finite size effects, and (iii) the strain amplitude on that 2-sphere, and the 
duration of GWs, should be large enough to generate field excitations.

\section{acknowledgements}
We are very grateful to Ajit M. Srivastava and Bharat Kumar for very useful discussions, 
suggestions, and comments on the manuscript. We also would like to thank A.P. Balachandran, 
T.R. Govindarajan, P.S. Saumia, Abhishek Atreya, Minati Biswal, and Subhroneel Chakrabarti 
for useful discussions and comments on the work.


\begin{thebibliography}{99}


\bibitem{FDM0} W. Hu, R. Barkana, and A. Gruzinov, Phys. Rev. Lett. {\bf 85}, 1158 (2000).

\bibitem{SFDM} B. Li, T. Rindler-Daller, and P.R. Shapiro, Phys. Rev. D {\bf 89}, 083536 (2014).

\bibitem{TRG} T.R. Govindarajan and N. Kalyanapuram, Mod. Phys. Lett. A {\bf 34}, 1950330 (2019). 

\bibitem{axionlike} L. Hui, J.P. Ostriker, S. Tremaine, and E. Witten, Phys. Rev. D {\bf 95}, 043541 (2017).

\bibitem{natureFDM} H.-Y. Schive, T. Chiueh, and T. Broadhurst, Nature Phys {\bf 10}, 496 (2014).

\bibitem{rev_superradiance} R. Brito, V. Cardoso, and P. Pani, {\it Superradiance: New Frontiers in 
Black Hole Physics}, 2nd ed., Lecture Notes in Physics Vol. {\bf 971} (Springer, Cham, 2020).

\bibitem{spr1} S.R. Dolan, Phys. Rev. D {\bf 87}, 124026 (2013).

\bibitem{spr2} R. Brito, V. Cardoso, and P. Pani, Class. Quantum Grav. {\bf 32}, 134001 (2015).

\bibitem{spr3} G. Ficarra, P. Pani, and H. Witek, Phys. Rev. D {\bf 99}, 104019 (2019).

\bibitem{spr4} D. Baumann, H.S. Chia, J. Stout, and L.t. Haar, JCAP {\bf 12}, 006 (2019).

\bibitem{spr_pulsar1} V. Cardoso, P. Pani and T.-T. Yu, Phys. Rev. D {\bf 95}, 124056 (2017).

\bibitem{spr_pulsar2} F.V. Day and J.I. McDonald, JCAP {\bf 10}, 051 (2019).

\bibitem{spr5} A. Arvanitaki and S. Dubovsky, Phys. Rev. D {\bf 83}, 044026 (2011).

\bibitem{spr_obsv1} A. Arvanitaki, M. Baryakhtar, and X. Huang, Phys. Rev. D {\bf 91}, 084011 (2015).

\bibitem{spr_obsv2} R. Brito, S. Ghosh, E. Barausse, E. Berti, V. Cardoso, I. Dvorkin, A. Klein, and 
P. Pani, Phys. Rev. D {\bf 96}, 064050 (2017).

\bibitem{spr_obsv3} O.A. Hannuksela, K.W.K. Wong, R. Brito, E. Berti, and T.G.F. Li, Nat. Astron. {\bf 3}, 447 (2019).

\bibitem{spr_obsv4} S. D'Antonio et al., Phys. Rev. D {\bf 98}, 103017 (2018).

\bibitem{spr_obsv5} M.J. Stott and D.J.E. Marsh, Phys. Rev. D {\bf 98}, 083006 (2018).

\bibitem{spr_obsv6} M. Isi, L. Sun, R. Brito, and A. Melatos, Phys. Rev. D {\bf 99}, 084042 (2019).

\bibitem{spr_obsv7} S. Ghosh, E. Berti, R. Brito, and M. Richartz, Phys. Rev. D {\bf 99}, 104030 (2019).

\bibitem{spr_obsv8} J. Zhang and H. Yang, Phys. Rev. D {\bf 99}, 064018 (2019).

\bibitem{spr_obsv9} K.K.Y. Ng, M. Isi, C.-J. Haster, and S. Vitale, Phys. Rev. D {\bf 102}, 083020 (2020).

\bibitem{spr_obsv10} K.K.Y. Ng, O.A. Hannuksela, S. Vitale, and T.G.F. Li, Phys. Rev. D {\bf 103}, 063010 (2021).

\bibitem{ULA_BBH} Q. Yang, L.-W. Ji, B. Hu, Z.-J. Cao, and R.-G. Cai, Res. Astron. Astrophys. {\bf 18}, 065 (2018).

\bibitem{sukanta} S. Choudhary, N. Sanchis-Gual, A. Gupta, J.C. Degollado, S. Bose, and J.A. Font,
Phys. Rev. D {\bf 103}, 044032 (2021).

\bibitem{sprReso1} D. Baumann, H.S. Chia, and R.A. Porto, Phys. Rev. D {\bf 99}, 044001 (2019).

\bibitem{sprReso2} E. Berti, R. Brito, C.F.B. Macedo, G. Raposo, and J.L. Rosa, Phys. Rev. D {\bf 99}, 104039 (2019).

\bibitem{sprReso3} D. Baumann, H.S. Chia, R.A. Porto, and J. Stout, Phys. Rev. D {\bf 101}, 083019 (2020). 

\bibitem{sprReso4}  M. Kavic, S.L. Liebling, M. Lippert, and J.H. Simonetti, JCAP {\bf 08}, 005 (2020). 

\bibitem{SMohanty} T.K. Poddar, S. Mohanty, and S. Jana, Phys. Rev. D {\bf 101}, 083007 (2020).

\bibitem{massiveScalar} B.C. Seymour and K. Yagi, Phys. Rev. D {\bf 102}, 104003 (2020).

\bibitem{numericalSim} B.J. Kavanagh, D.A. Nichols, G. Bertone, and D. Gaggero, Phys. Rev. D {\bf 102}, 083006 (2020).

\bibitem{axion_search} J. Huang, M.C. Johnson, L. Sagunski, M. Sakellariadou, and J. Zhang, Phys. Rev. D {\bf 99}, 063013 (2019).

\bibitem{rev_alp_mrgr1} G. Bertone et al., SciPost Phys. Core {\bf 3}, 007 (2020). 

\bibitem{rev_alp_mrgr2} J.-F. Fortin et al., Int. J. Mod. Phys. D {\bf 30}, 2130002 (2021).

\bibitem{our1} S.S. Dave and S. Digal, Phys. Rev. D {\bf 103}, 116007 (2021).

\bibitem{string_axiverse} A. Arvanitaki, S. Dimopoulos, S. Dubovsky, N. Kaloper, and J. March-Russell, 
Phys. Rev. D {\bf 81}, 123530 (2010).

\bibitem{review_ULA} E.G.M. Ferreira, Astron. Astrophys. Rev. {\bf 29}, 7 (2021). 

\bibitem{axn1} J. Preskill, M.B. Wise, and F. Wilczek, Phys. Lett. B {\bf 120}, 127 (1983). 

\bibitem{axn2} L.F. Abbott and P. Sikivie, Phys. Lett. B {\bf 120}, 133 (1983).

\bibitem{axn3} M. Dine and W. Fischler, Phys. Lett. B {\bf 120}, 137 (1983).

\bibitem{carroll} S. Carroll, {\it Spacetime and Geometry; An Introduction to General 
Relativity} (Addison Wesley, Reading, MA, 2004).

\bibitem{GWbbh} B.P. Abbott et al. (LIGO Scientific and Virgo Collaborations), Phys. Rev. Lett. 
{\bf 116}, 061102 (2016).

\bibitem{leapfrog} W.H. Press, S.A. Teukolsky, W.T. Vetterling, and B.P. Flannery, {\it Numerical Recipes
in Fortran 77; The Art of Scientific Computing}, 2nd ed. Vol. 1 (Syndicate, New York and Melbourne, 1997). 

\bibitem{waveform1} L. Blanchet, C. R. Physique {\bf 20}, 507 (2019).

\bibitem{sg1} B.P. Abbott et al., Phys. Rev. D {\bf 80}, 102001 (2009).

\bibitem{sg2} J. Abadie et al., Phys. Rev. D {\bf 81}, 102001 (2010).

\bibitem{sg3} J. Abadie et al., Phys. Rev. D {\bf 85}, 122007 (2012).

\bibitem{sg4} B.P. Abbott et al., Phys. Rev. D {\bf 94}, 102001 (2016). 

\bibitem{sg5} B.P. Abbott et al., Phys. Rev. D {\bf 101}, 084002 (2020).

\bibitem{CCSN1} K. Kotake, C. R. Physique {\bf 14}, 318 (2013).  


\end{thebibliography}
\end{document}